\documentclass[12pt]{article}
\usepackage{epsf,epsfig,amsmath}
\begin{document}
\title{Comment on Papers by Gill and Gill, Weihs, Zeilinger and Zukowski}

\author{Karl Hess$^1$ and Walter Philipp$^2$}

\date{$^1$ Beckman Institute, Department of Electrical Engineering
and Department of Physics,University of Illinois, Urbana, Il 61801
\\ $^{2}$ Beckman Institute, Department of Statistics and Department of
Mathematics, University of Illinois,Urbana, Il 61801 \\ }
\maketitle

\begin{abstract}

We show that the proofs of Gill as well as of Gill, Weihs ,
Zeilinger and Zukowski contain serious mathematical and physical
deficiencies which render them invalid.

\end{abstract}

The purpose of this note is to point out several mathematical
errors in the papers by Gill \cite{mang} and by Gill, Weihs,
Zeilinger and Zukowski (GWZZ) \cite{gwzz} on their mathematical
model for EPR experiments. In the class of EPR experiments
considered here, correlated pairs (e.g. a pair of electrons in the
singlet state) are emitted from a source $S$ to two spatially
separated analyzer stations $S_1$ and $S_2$. Each particle of a
given pair carries the same information package $\Lambda$. It is
assumed that $\Lambda$ is a random variable, defined on some
probability space $\Omega$, whose elements are denoted by
$\omega$. One can think of the $\omega$'s as being simple,
indecomposable experiments in the sense of Feller. For instance
$\omega$ can be thought as being the experiment of sending out a
correlated pair from $S$, as we will assume in the present note,
or as being the experiment of sending out $\lambda$ (a value that
the random variable $\Lambda$ assumes) which can be treated in the
same way. The measuring instruments in the analyzer stations $S_1$
and $S_2$ are represented by two unit vectors each, $\bf a$ and
$\bf d$ in $S_1$ and $\bf b$ and $\bf c$ in $S_2$. The
measurements are denoted by $A_{\bf a}(\Lambda(\omega)), A_{\bf
d}(\Lambda(\omega))$ in $S_1$ and by $B_{\bf b}(\Lambda(\omega))$
and $B_{\bf c}(\Lambda(\omega))$ in $S_2$. The $A$'s and the $B$'s
are considered to be random variables, defined on $\Omega$, and to
be assuming the values +1 and -1, only.

GWZZ \cite{gwzz} introduce the entity (their Eq. (5), see also
Gill \cite{gbs})
\begin{equation}
\Delta = 1\{\omega:A_{\bf a}(\Lambda(\omega)) = B_{\bf c}(
\Lambda(\omega))\} - 1\{\omega:A_{\bf a}(\Lambda(\omega)) = B_{\bf
b}( \Lambda(\omega))\}\nonumber
\end{equation}
\begin{equation}\text{                          }
 - 1\{\omega:A_{\bf
d}(\Lambda(\omega)) = B_{\bf b}( \Lambda(\omega))\} -
1\{\omega:A_{\bf d}(\Lambda(\omega)) = B_{\bf c}(
\Lambda(\omega))\}\label{mc1}
\end{equation}
where $1\{...\}$ denotes the indicator of the event in curly
brackets. The indicator equals 1 if the equality holds (i.e.
$\omega$ belongs to the event in question) and equals 0 otherwise.
GWZZ claim that the $\Delta$ introduced in Eq.(\ref{mc1}) assumes
only the values 0 and -2. Clauser, Horne, Shimony and Holt (CHSH)
introduced the by now classical entity
\begin{equation}
\Gamma:= A_{\bf a}.B_{\bf b} - A_{\bf a}.B_{\bf c} - A_{\bf
d}.B_{\bf b} - A_{\bf d}.B_{\bf c}\label{mc2}
\end{equation}
They showed by a simple arithmetic manipulation involving the
factoring of $A_{\bf a}$ and $A_{\bf d}$ respectively that $\Gamma
= 2$ or $-2$. Now obviously we have e.g.
\begin{equation}
1\{A_{\bf a} = B_{\bf b}\} = (A_{\bf a}.B_{\bf b} + 1)/2 \nonumber
\end{equation}
plus three similar relations. Substituting these into
Eq.(\ref{mc1}) we obtain
\begin{equation}
\Delta = \Gamma/2 - 1 \label{mc3}
\end{equation}
and therefore $\Delta = 0$ or $-2$. Thus the GWZZ approach is a
trivial modification of the CHSH approach.

The claim by GWZZ that $\Delta = 0$ or $-2$ as well as the claim
by CHSH that $\Gamma = 2$ or $-2$ are based on the assumption that
each $A$ with the same setting is the same (either $+1$ or $-1$)
independent of what the setting for $B$ is. At first glance, this
assumption appears to follow from the definition of locality.
However, locality can only demand that $A$, as a function, be
independent of the setting chosen for the given pair in station
$S_2$. $A$ does not have to be the same in the events $\{\omega: A
= B\}$, respectively in the products $A.B$, when we have the same
setting in $S_1$ and different settings in $S_2$. We demonstrate
this error in two steps, the first involving basic physics and
probability theory, the second involving more detailed physics and
the hypothesis of certain time dependencies of EPR experiments.

On grounds of basic physics and elementary probability theory, it
is impossible to have at any given measurement time-period for any
given correlated pair two (or more) different macroscopic
configurations or settings of instruments that are chosen and
observed by the experimenter. It is also physically impossible to
generate two or more guaranteed identical correlated pairs and to
send them simultaneously to several different stations with
different settings. Therefore, no experiment $\omega$ can be
performed that can simultaneously determine the values of all four
indicators in Eq.(\ref{mc1}). Each experiment $\omega$ can
determine only the value of one indicator, the other three are
counterfactual. Four separate experiments are needed to determine
the values of all four indicators, as can also be seen from the
experimental arrangement of GWZZ \cite{gwzz}. Hence $\Delta$
corresponds to a non-performable experiment and $\Delta$ is a
non-computable entity. Thus the definition of $\Delta$ in
Eq(\ref{mc1}) is meaningless because $\Delta$ is not a
well-defined function on $\Omega$. The mistake that GWZZ make
arises from their Fig.2 that cannot be reconciled with the
physical facts of their actual EPR experiment and the
corresponding appropriate construction of a sample space
\cite{feller}.

For the reasons just given, the $\omega$'s in each of the four
different events of Eq(\ref{mc1}) must be all different and hence
there is no guarantee that the corresponding values of $\Lambda$
are the same. Thus there is no guarantee that the values of the
various $A$'s and $B$'s are the same although they are tagged by
the the same vectors ${\bf a},{\bf d}$ and ${\bf b},{\bf c}$
respectively. As a consequence the arithmetic manipulation
mentioned above cannot be carried out. Hence, at this point of
GWZZ's proof \cite{gwzz}, the assertion that $\Delta$ assumes only
the values 0 and -2 is false.

Because $\Delta$ is not a function on $\Omega$, $\Delta$ is not a
random variable. Thus it makes no sense to take conditional
expectations of copies of $\Delta$ with respect to past sigma
fields. Thus and because the assertion that $\Delta = 0$ or $-2$
is false it makes even less sense to speak in this context about
supermartingales and to apply exponential probability bounds for
supermartingales, as was done by Gill \cite{mang}, nor does it
make sense to apply the strong law of large numbers to the
sequence of independent copies of $\Delta$ because $\Delta$ is not
a random variable.

There is an ironic twist to the logic in \cite{mang} and
\cite{gwzz}. If the authors are that convinced that $\Delta$ of
Eq(\ref{mc1}) assumes only the values 0 and -2, what is the
purpose of considering a large number of these non-performable
experiments based on $\Delta$? What is the purpose of then
averaging the results of these non-performable experiments and
then to show that these averages contradict the results of actual
experiments when in fact {\it EACH} of these non-performable
experiments is already predicted by Eq(\ref{mc1}) to yield the
outcome 0 or -2? All that is needed is a one on one comparison of
the results of a few of these non-performable experiments with the
results of actual experiments.

Having seen that GWZZ \cite{gwzz} and Gill \cite{mang} violate the
syntax of probability theory as outlined e.g. in Feller's
\cite{feller} work, we proceed to ask the question whether the
proof of GWZZ can be saved by certain assumptions on the parameter
space? Indeed, if $\Lambda$ assumes only one value e.g.
$\lambda_1$, then Eq(\ref{mc1}) is obviously validated. More
generally, the procedure of GWZZ can be validated if and only if
any set of data (outcomes) can be {\it reordered} in rows
corresponding to Eq(\ref{mc1}) give or take a few left over terms
that are insignificant for large numbers of experiments. The proof
of this statement is trivial. We have shown, however, in past
publications \cite{mer} that such reordering is limited to
parameters $\Lambda$ having a probability distribution with
countable support and certain restricted time dependencies. As can
easily be seen, the Bell type proofs will not go forward if, for
instance, the $\lambda$'s are replaced by the necessarily all
different clock-times of measurement. We have also shown that the
physically very reasonable extension of the parameter space to
include setting and time dependent instrument parameters puts Bell
type proofs and reordering arguments to a halt \cite{mer},
\cite{eur}. Instrument parameters are not subject to any further
locality conditions and restrictions because they describe the
instruments and therefore are permitted to depend on the setting
of the given instrument. It is also physically reasonable to let
these setting dependent instrument parameters depend on a
clock-time of the measurement. This time is, as mentioned above,
guaranteed to be different for different settings. The
introduction of such physically reasonable time dependencies
invalidates, a fortiori, the assumptions that lead to the outcome
0 or -2 for $\Delta$.

As a final remark, we would like to point out that the concept of
``time" used by Gill is incommensurate with the time-concept that
has evolved after Zeno and the early Greeks. Gill \cite{mang}
counts as history and time only the past randomly chosen settings
and the past measurements of $A$ and $B$. Time thus emerges as
representing a finite number of events that have very specific
properties. This primitive notion of time is at the basis of
Gill's conditioning on past sigma fields and his false claim that
the structure he considers is a supermartingale. Paradoxical
consequences of such oversimplified views are not surprising. It
is known since Zeno that the strangest paradoxes arise from
linking time to a countable number of specific events. Gill's
oversimplified model and time-concept can not be used to explain
the motion of Foucault's pendulum or a compass based on
gyroscopes; nor should it be used to explain correlated spin pairs
for that matter. In summary, GWZZ's \cite{gwzz} and Gill's
\cite{mang} proofs are physically overly simplified and
mathematically artificial, negligent, incorrect and circular;
particularly when considering the highest standards that must be
applied when the foundations of scientific frameworks are
investigated.


\begin{thebibliography}{99}

\bibitem{mang} R. D. Gill, quant-ph/0110137 (2001), Final version
18. September 2002.

\bibitem{gwzz} R. D. Gill, G. Weihs, A.Zeilinger and M. Zukowski,
quant-ph/0204169, April 30 (2002); see also R. D. Gill, G. Weihs,
A. Zeilinger and M. Zukowskyi, Proc. Nat. Academy of Sci. (USA),
Vol. 99, 14632-14635 (2002)

\bibitem{gbs} R. D. Gill, Proc. Conference, Foundations of
Probability and Physics - 2, Vaxjo, Sweden, p189 (2002)

\bibitem{feller} W. Feller, ``An Introduction to Probability
Theory and its Applications" Vol 1, 3rd edition, Wiley Series in
Probability and Mathematical Statistics, pp1-9 (1968)

\bibitem{mer} K. Hess and W. Philipp, quant-ph/0305037 (2003)

\bibitem{eur} K. Hess and W. Philipp, Europhys. Lett. Vol. 57,
775-781 (2002)





\end{thebibliography}
\end{document}